\documentclass{iau}
\usepackage{graphicx}
\usepackage{subfigure}
\usepackage{amsmath}
\usepackage{amssymb,amsfonts}
\usepackage{bm}

\usepackage{url}
\usepackage{amssymb} 
\usepackage{amsfonts}
\DeclareMathOperator{\prox}{prox}

\title[JD 11.~~Pre-solar grains and AGB stars] 
{PRISM: Sparse recovery of the primordial spectrum from WMAP9 and Planck datasets}

\author[P. Paykari \etal\ ]   
{P. Paykari$^1$, F. Lanusse$^1$,  J.-L. Starck$^1$,  F. Sureau$^1$ \and  J. Bobin$^1$}

\affiliation{$^1$Service d'Astrophysique, CEA Saclay, F-91191 Gif sur Yvette cedex, France. \\
email: {\tt paniez.paykari@cea.fr}}

\pubyear{2014}
\volume{306}  
\pagerange{119--126}
\jname{Statistical Challenges in 21st Century Cosmology}
\editors{A.C. Editor, B.D. Editor \& C.E. Editor, eds.}
\begin{document}

\maketitle

\begin{abstract}
The primordial power spectrum is an indirect probe of inflation or other structure-formation mechanisms. We introduce a new method, named \textbf{PRISM}, to estimate this spectrum from the empirical cosmic microwave
background (CMB) power spectrum. This is a sparsity-based inversion method, which leverages a sparsity prior on features in the primordial spectrum in a wavelet dictionary to regularise the inverse problem. This non-parametric approach is able to reconstruct the global shape as well as localised features of spectrum accurately and proves to be robust for detecting deviations from the currently favoured scale-invariant spectrum.
We investigate the strength of this method on a set of WMAP nine-year simulated data for three types of primordial spectra and then process the WMAP nine-year data as well as the Planck PR1 data. We find no significant departures from a near scale-invariant spectrum. 
\keywords{\small{Cosmology: Primordial Power Spectrum, Methods: Data Analysis, Statistical}}
\end{abstract}

\firstsection 

\section{Introduction}
\label{sec:intro}

Inflation \cite[(Guth 1981,]{guth} \cite[Linde 1982)]{Linde-inflation1982} is currently the most favoured model describing the early Universe, in which perturbations are produced by quantum fluctuations during the epoch of an accelerated expansion. The simplest models of inflation predict a 
near scale-invariant primordial power spectrum described in terms of a spectral index $n_s$, an amplitude of the perturbations $A_s$ and a possible `running' $\alpha_s$ of the spectral index
\begin{equation}
\small{
P(k)=A_s \left(\frac{k}{k_p}\right)^{n_s-1+\frac{1}{2}\alpha_s\ln \left({k}/{k_p}\right)}\;,
}
\end{equation}
where $k_p$ is a pivot scale. The near scale-invariant spectrum with $n_s<1$ fits the current observations very well 
\cite[(Ade \etal\ 2013a,]{PlanckCP} \cite[Ade \etal\ 2013b)]{PlanckPk}.

The recent Planck data, combined with the WMAP large-scale polarisation, constrain the spectral index to $n_{s}=0.9603 \pm0.0073$ 
\cite[(Ade \etal\ 2013a)]{PlanckCP}, ruling out exact scale invariance at over $5 \sigma$. Also, Planck does not find a statistically significant running of the scalar spectral index, obtaining $\alpha_s =-0.0134\pm0.0090$. In addition, high-resolution CMB experiments, such as the South Pole Telescope (SPT)\footnote{\url{http://pole.uchicago.edu/spt/index.php}}, report a small running of the spectral index; $-0.046 < \alpha_s < -0.003$ at $95\%$ confidence 
\cite[(Hou \etal\ 2012)]{SPT_cos}. However, in general, any such detections have been small and consistent with zero.  

Determining the shape of the primordial spectrum generally consists of two approaches, one by parametrisation and the second by reconstruction. Non-parametric methods suffer from the non-invertibility of the transfer function that descries the {\it transfer} from $P(k)$ to the CMB spectrum:
$C_{\ell}^{\textrm{th}}=4\pi\int_{0}^{\infty}d\ln k\Delta_{\ell}^{2}(k)P(k)$,
where $\ell$ is the angular wavenumber and $\Delta_{\ell}(k)$ is the radiation transfer function. Because of the singularity of the transfer function and the limitations on the data due to effects such as projection, cosmic variance, instrumental noise, point sources, and etc., a sensitive algorithm is necessary for an accurate reconstruction of the primordial power spectrum from CMB data. 

\section{PRISM Algorithm}
\label{sec:prism}

A CMB experiment measures the anisotropies in the CMB temperature $\Theta(\vec{p})$ in direction $\vec{p}$, which is described as $T(\vec{p}) = T_{\mathrm{CMB}} [1 + \Theta(\vec{p}) ]$. 
This field can be expanded in terms of spherical harmonic functions $Y_{\ell m}$ as
$\Theta(\vec{p}) = \sum_\ell \sum_m a_{\ell m} Y_{\ell m}(\vec{p}),$
where $a_{\ell m}$ are the spherical harmonic coefficients, that have a  Gaussian distribution with zero mean, $\langle a_{\ell m} \rangle = 0$, and variance $ \langle a_{\ell m} a^*_{\ell^{\prime} m^{\prime}} \rangle = \delta_{\ell \ell^{\prime}} \delta_{mm^{\prime}} C_\ell^{\textrm{th}}$. In practice, we are restricted by cosmic variance, additive instrumental noise and also, due to the different Galactic foregrounds, we need to mask the observed CMB map, which induces correlations between different modes. Taking these effects into account and following the MASTER method from 
\cite[Hivon \etal\ (2002)]{MASTER}, the pseudo power spectrum $\widetilde{C}_\ell$ and the empirical power spectrum $\widehat{C}^{\mathrm{th}}_\ell$, which is defined as $\widehat{C}^{\mathrm{th}}_\ell = 1/(2\ell + 1) \sum_m | a_{\ell m} |^2$, can be related through their ensemble averages
$
\langle \widetilde{C}_\ell \rangle = \sum_{\ell^\prime} M_{\ell \ell^\prime}   \langle \widehat{C}^{\mathrm{th}}_{\ell^\prime} \rangle +  \langle \widetilde{N}_\ell \rangle
$,
where $M_{\ell \ell^\prime}$ describes the mode-mode coupling due to the mask. We note that in this expression $ \langle \widehat{C}^{\mathrm{th}}_{\ell^\prime} \rangle =  C^{\mathrm{th}}_{\ell^\prime} $, and we set $C_\ell = \langle \widetilde{C}_\ell \rangle$ and $N_\ell = \langle \widetilde{N}_\ell \rangle$,
where $C_\ell$ and $N_\ell$ refer to the CMB and the noise spectra of the masked maps, respectively.
We assume the pseudo spectrum $\widetilde{C}_\ell$ follows a $\chi^2$ distribution with $2 \ell + 1$ degrees of freedom and can be modelled as
$
\widetilde{C}_\ell =\left( \sum_{\ell^\prime} M_{\ell \ell^\prime} C^{\mathrm{th}}_{\ell^\prime} + N_\ell \right) Z_\ell \label{eq:pseudoToTheoPS}
$,
where $Z_\ell$ is a random variable representing a multiplicative noise.

The relation between the discretized primordial spectrum $P_{k}$ and the pseudo spectrum $\widetilde{C}_\ell$, computed on a masked noisy map of the sky, can be condensed in the following form
\begin{equation}
\small{
\widetilde{C}_\ell = \left( \sum_{\ell^\prime k } M_{\ell \ell^\prime} T_{\ell^\prime k}P_{k} +   N_\ell \right) Z_\ell\;,
\label{eq:complete-problem-mult}
}
\end{equation}
where $T_{\ell k}=4 \pi \Delta\ln k\,\Delta_{\ell k}^2$. Because of the non-invertibility of $\mathbf{T}$, recovering $P_k$ constitutes an ill-posed inverse problem. This inverse problem can be regularised in a robust way by using the sparse nature of the reconstructed signal as a prior. Furthermore, sparse recovery has already been successfully used in the TOUSI algorithm 
\cite[(Paykari \etal\ 2012)]{Tousi} 
to handle the multiplicative noise term, where after the stabilisation, the noise can be treated as an additive Gaussian noise with zero mean and unit variance. 
The sparse regularisation framework means that if $P_k$ can be sparsely represented in an adapted dictionary $\mathbf{\Phi}$, then this problem, known as the basis pursuit denoising BPDN, can be recast as an optimisation problem, formulated as
\begin{equation}
\small{
\min\limits_X \frac{1}{2} \parallel C_\ell - (\mathbf{M}\mathbf{T} X + N_\ell) \parallel_2^2 + \lambda \parallel \mathbf{\Phi}^t X \parallel_0\;,
\label{eq:bp-lagragian}
}
\end{equation}
where $X$ is the reconstructed estimate for $P_k$. The first term imposes a $\ell_2$ fidelity constraint to the data while the second term promotes the sparsity constraint of the solution in dictionary $\mathbf{\Phi}$, by tuning $\lambda$. The $\ell_0$ optimisation problem in the second term cannot be solved directly, therefore, we estimate the solution by solving a sequence of relaxed problems using a re-weighted $\ell_1$ minimisation technique 
\cite[(Candes 2007)]{Candes2007}
\begin{equation}
\small{
\min\limits_X \frac{1}{2} \parallel \frac{1}{\sigma_\ell}\overline{R}_\ell(X) \parallel_2^2 + K \sum_i \lambda_i | [ \mathbf{W} \mathbf{\Phi}^t  X ]_i  |\;,
\label{eq:reweighted-bp-lagragian}
}
\end{equation}
where $\mathbf{W}$ is a diagonal matrix applying a different weight for each wavelet coefficient, $\overline{R}_\ell(X)$ is our estimate of the residual in the fidelity term in Eq.~\eqref{eq:bp-lagragian} and $\sigma_\ell$ is the noise on the residual. Our specific choice of the $\lambda_i$ has allowed us to use a single regularisation parameter $K$ which translates into a significance level threshold for feature detection. The relaxed problem \eqref{eq:reweighted-bp-lagragian} can be solved by ISTA (iterative soft-thresholding algorithm)
\begin{align}
\widetilde{X}^{n+1} &= X^{n} + \mu \mathbf{T}^{t} \mathbf{M}^{t} \frac{1}{\sigma_\ell}\overline{R}_\ell(X^{n})\;, \\
X^{n+1} &= \prox_{K  \mu \parallel \lambda \odot W \Phi^t \cdot \parallel_1 } \left(\widetilde{X}^{n+1} \right) \;,
\label{eq:ISTA}
\end{align}
where $\mu$ is an adapted step size and $\prox_{K  \mu \parallel \lambda \odot W \Phi^t \cdot \parallel_1 }$ is the proximal operator corresponding to the sparsity constraint. Full details of PRISM is in 
\cite[Paykari \etal\ (2014)]{PRISM_WMAP9}.

\section{Results and Conclusion}
\label{sec:results}

\begin{figure}[ht]
\begin{center}
\begin{tabular}{cc}
\subfigure{\includegraphics[width=0.51\textwidth]{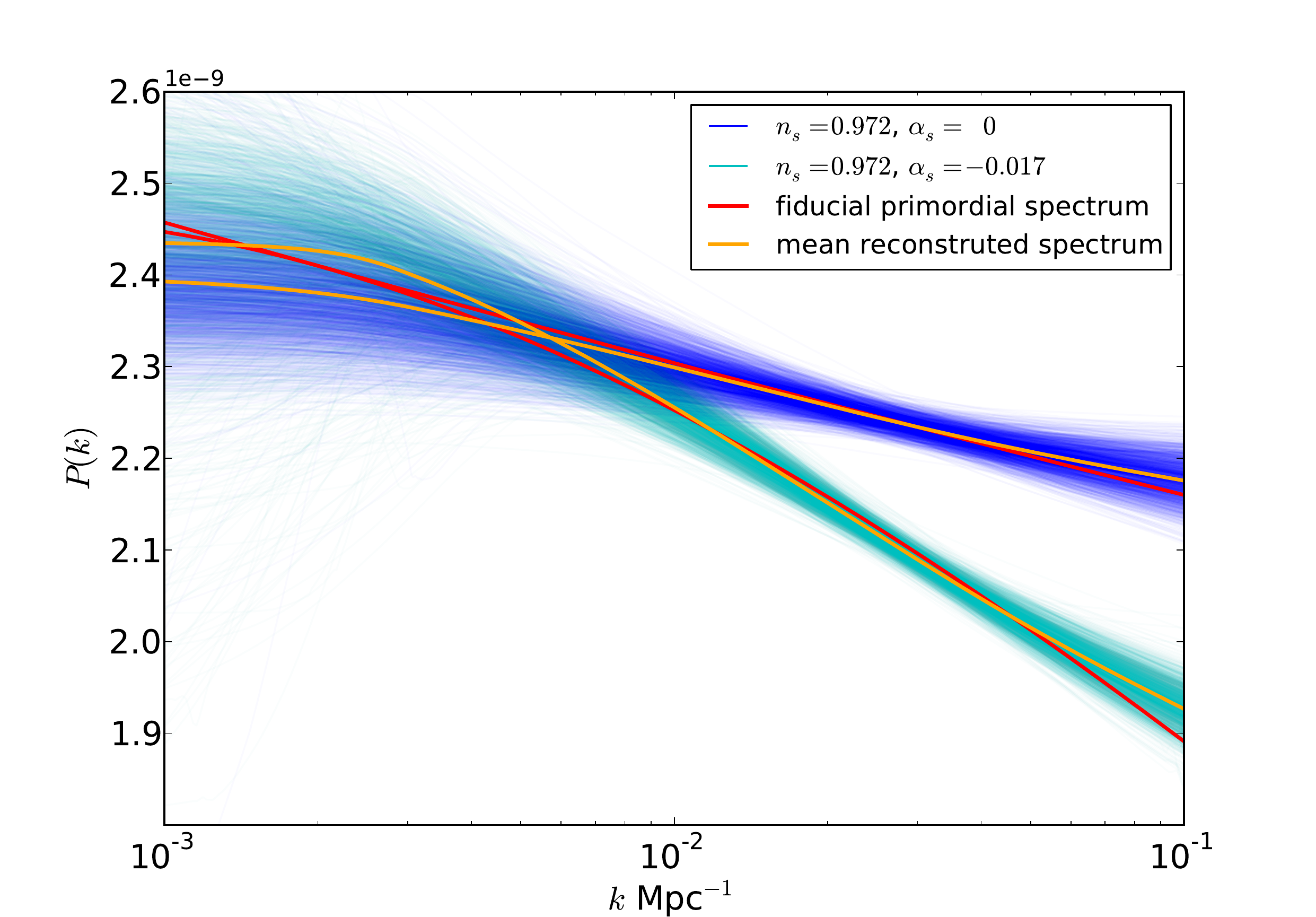}}  & 
\subfigure{\includegraphics[width=0.51\textwidth]{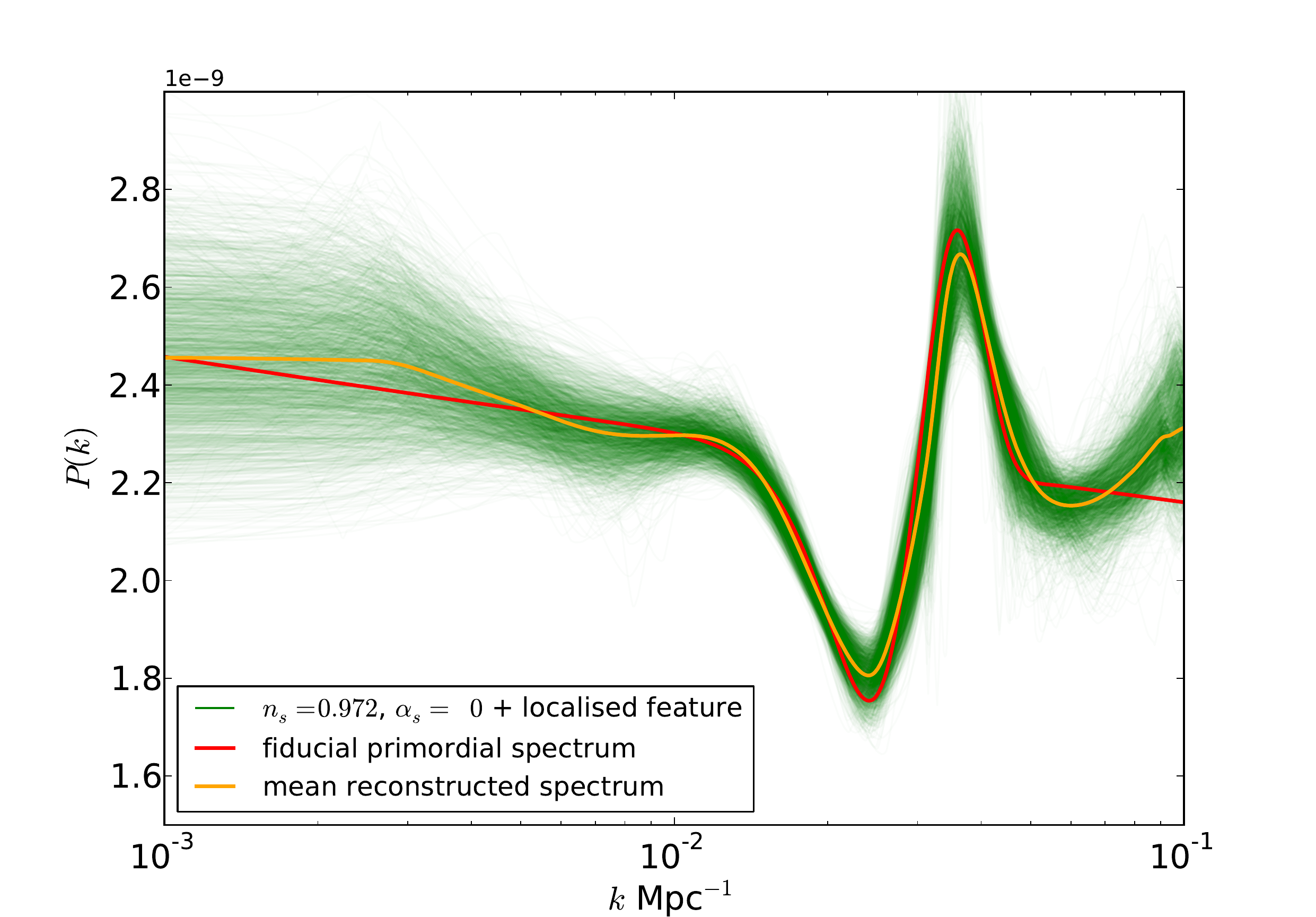}}  
\end{tabular}
\caption{Reconstructions for three different types of primordial spectra: near scale-invariant spectrum, spectrum with a small running, and spectrum with a localised feature. Left: blue lines show the 2000 reconstructed spectra with $n_s=0.972$ and $\alpha_s=0$ and the cyan lines show the 2000 reconstructions for $n_s=0.972$ and $\alpha_s=-0.017$. As can be seen, for $k>0.015 \; \mathrm{Mpc}^{-1}$ PRISM can reconstruct the primordial spectra with such accuracy that the two are easily distinguishable. Right: the 2000 reconstructions of a spectrum with $n_s=0.972$, $\alpha_s=0.0$, and an additional feature around $k=0.03$ Mpc$^{-1}$ is shown in green. As can be seen, PRISM is able to recover both the position and the amplitude of the feature with great accuracy. In all cases the mean of the reconstructions is shown in orange and the fiducial input spectrum in red.}
\label{fig:sim}
\end{center}
\end{figure}

\begin{figure}[ht]
\begin{center}
\begin{tabular}{cc}
\subfigure{\includegraphics[width=0.51\textwidth]{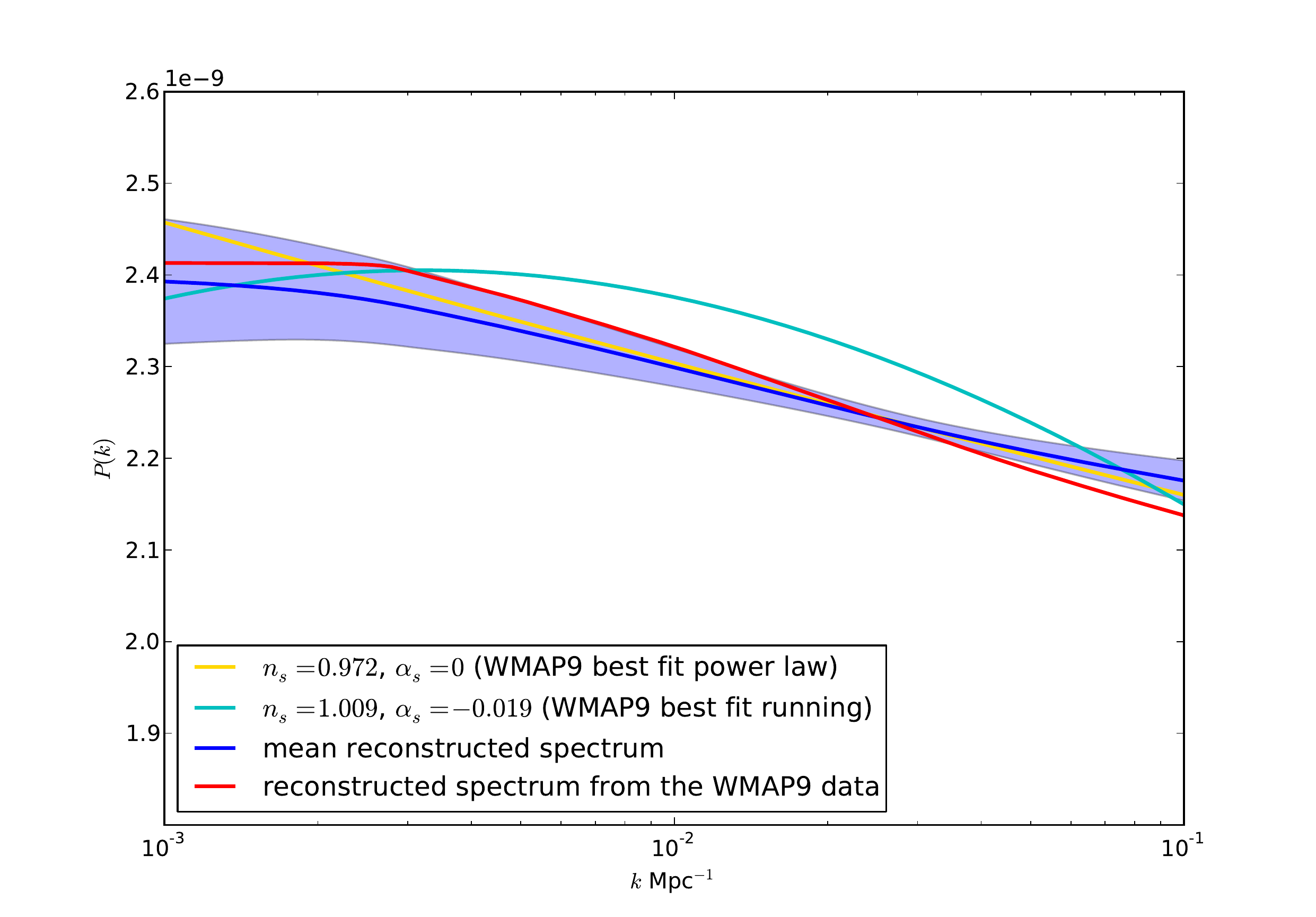}}  & 
\subfigure{\includegraphics[width=0.43\textwidth]{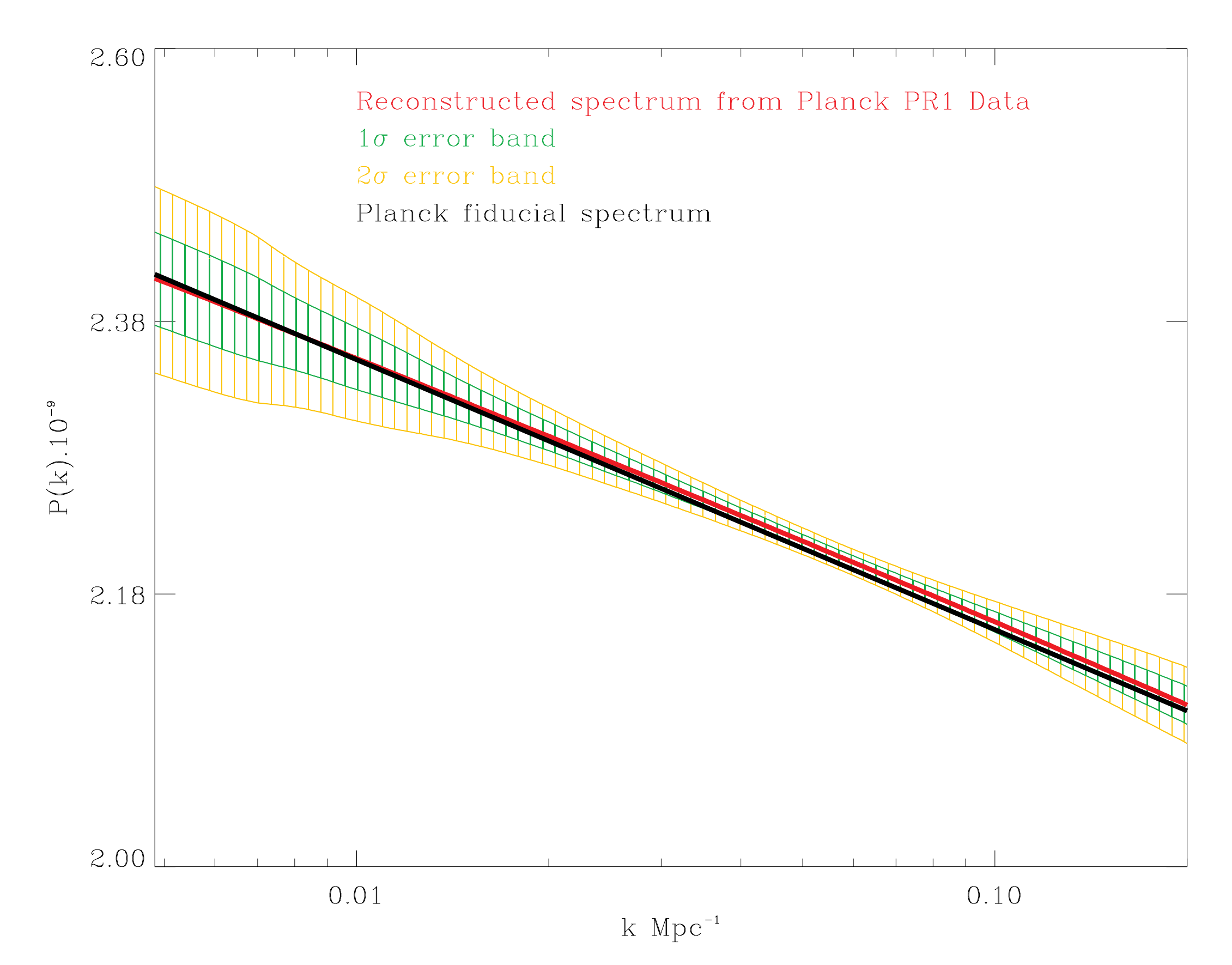}}  
\end{tabular}
\caption{Left: Reconstruction of the primordial spectrum from the WMAP nine-year spectrum is shown in red. The mean of the reconstructions for $n_s=0.972$ and $\alpha_s = 0$ is shown in solid dark blue line, with the $1\sigma$ interval around the mean shown as a shaded blue region. The WMAP nine-year fiducial primordial spectrum with $n_s=0.972$ and $\alpha_s = 0$ is shown in yellow and in cyan we show the best-fit primordial spectrum with a running with $n_s=1.009$ and $\alpha_s = -0.019$. Right: Reconstructed primordial spectrum from Planck PR1 data is shown in red. The $1\sigma$ and $2\sigma$ errors are also shown as green and yellow bands respectively. The fiducial primordial power spectrum with $n_s=0.9626$ is shown in black solid line. We note that the error bands do not include the errors due to point sources and beam uncertainties. We have not detected any significant deviations, whether local or global, from a scale-invariant spectrum.}
\label{fig:data}
\end{center}
\end{figure}

We apply PRISM to $2000$ simulated pseudo spectra for three different types of $P_k$. The reconstructions are shown in Fig.~\ref{fig:sim} with the details being explained in the caption. We also apply PRISM to the WMAP nine-year and Planck PR1 LGMCA CMB pseudo spectra \cite[(Bobin \etal\  2014)]{PR1_WPR1} and the results are shown in Fig.~\ref{fig:data}. We have not detected any significant deviations, whether local or global, from a scale-invariant primordial spectrum.

The reconstruction of the primordial spectrum is limited by different effects on different scales. On very large scales, there are fundamental physical limitations due to the cosmic variance and the more severe geometrical projection of the modes, meaning the primordial spectrum cannot be fully recovered on these scales, even in a perfect CMB measurement. On small scales we are limited by the instrumental noise, point sources, beam uncertainties, and etc.. This leaves us with a window through which we can recover the primordial spectrum with a good accuracy. Nevertheless, as can be seen in the left plot of Fig.~\ref{fig:sim}, for $k>0.015 \;\mathrm{Mpc}^{-1}$, PRISM can easily distinguish the two very similar spectra and performs very well in reconstructing the position and amplitude of the featured spectra, as shown on the right hand plot. 

To conclude, the PRISM algorithm uses the sparsity of the primordial spectrum as well as an adapted modelling of the noise of the CMB spectrum to recover the primordial spectrum. This algorithm assumes no prior shape for the primordial spectrum and does not require a coarse binning of the primordial spectrum, making it sensitive to both global smooth features (e.g. running of the spectral index) as well as local sharp features (e.g. a bump or an oscillatory feature). Another advantage of this method is that the regularisation parameter can be specified in terms of a signal-to-noise significance level for feature detection. These advantages make this technique very suitable for investigating different types of departures from scale invariance, whether it is the running of the spectral index or some localised sharp features as predicted by some of the inflationary models. We have applied PRISM to LGMCA WMAP nine-year and Planck PR1 spectra and have not detected any significant deviations from a scale-invariant power spectrum, whether local or global such as a running of the spectral index.

The developed C++ and IDL codes is released in iSAP (Interactive Sparse astronomical data Analysis Packages) via the web site
\url{http://cosmostat.org/isap.html}.
 

\end{document}